\documentclass[twocolumn,superscriptaddress,showpacs,prl]{revtex4}

\usepackage[dvips]{graphicx}% Include figure files
\usepackage{dcolumn}% Align table columns on decimal point
\usepackage{bm}% bold math

\begin{document}

\title{Fermi Surface and Order Parameter Driven Vortex Lattice Structure Transitions in Twin-Free YBa$_{2}$Cu$_{3}$O$_{7}$}

\author{J.S.~White}
\affiliation{School of Physics and Astronomy, University of Birmingham, Edgbaston, Birmingham, B15 2TT, UK.}
\author{V.~Hinkov}
\affiliation{Max Planck Institut f\"{u}r Festk\"{o}rperforschung, D-70569 Stuttgart, Germany.}
\author{R.W.~Heslop}
\author{R.J.~Lycett}
\author{E.M.~Forgan}
\author{C.~Bowell}
\affiliation{School of Physics and Astronomy, University of Birmingham, Edgbaston, Birmingham, B15 2TT, UK.}
\author{S.~Str\"{a}ssle}
\affiliation{Physik-Institut der Universit\"{a}t Z\"{u}rich, CH-8057 Z\"{u}rich, Switzerland.}
\author{A.B.~Abrahamsen}
\affiliation{Ris\o~National Laboratory for Sustainable Energy, DTU, DK-4000 Roskilde, Denmark.}
\author{M.~Laver}
\affiliation{NIST Center for Neutron Research, Gaithersburg, Maryland 20899, USA.}
\author{C.D.~Dewhurst}
\affiliation{Institut Laue-Langevin, 6 rue Jules Horowitz, 38042 Grenoble, France.}
\author{J.~Kohlbrecher}
\author{J.L.~Gavilano}
\author{J.~Mesot}
\affiliation{Paul Scherrer Institute, ETH Z\"{u}rich and EPF Lausanne, Villigen PSI, CH-5232, Switzerland.}
\author{B.~Keimer}
\affiliation{Max Planck Institut f\"{u}r Festk\"{o}rperforschung, D-70569 Stuttgart, Germany.}
\author{A.~Erb}
\affiliation{Walther Meissner Institut, BAdW, D-85748 Garching, Germany.}

\date{\today}

\begin{abstract}

%We report on small-angle neutron scattering (SANS) measurements of the magnetic vortex lattice (VL) in fully oxygenated YBa$_2$Cu$_3$O$_7$. In our detwinned sample, pinning to twin-domain boundaries along $\{110\}$ was largely suppressed, allowing the study of the intrinsic VL structure at 2~K, in fields of up to 10.8~T. At intermediate fields, we observe a distorted hexagonal VL structure which eluded previous identification since it is not pinned to the $\{110\}$ planes. This phase is separated from a low-field hexagonal phase of different orientation and distortion by a first-order transition at 2.0(2)~T, which is probably driven by Fermi-surface effects. We observe another first-order transition at 6.7(2)~T to a rhombic structure with a distortion of opposite sign. We argue that this high-field transition marks a crossover from a regime where Fermi surface anisotropy is dominant, to one where the VL structure and distortion is controlled by the superconducting order parameter.

We report on small-angle neutron scattering studies of the intrinsic vortex lattice (VL) structure in detwinned YBa$_2$Cu$_3$O$_7$ at 2~K, and in fields up to 10.8~T. Due to the suppressed pinning to twin-domain boundaries, a new distorted hexagonal VL structure phase is stabilized at intermediate fields. It is separated from a low-field hexagonal phase of different orientation and distortion by a first-order transition at 2.0(2)~T that is probably driven by Fermi surface effects. We argue that another first-order transition at 6.7(2)~T, into a rhombic structure  with a distortion of opposite sign, marks a crossover from a regime where Fermi surface anisotropy is dominant, to one where the VL structure and distortion is controlled by the order parameter anisotropy.

\end{abstract}


% PACS, the Physics and Astronomy Classification Scheme.
%Use showkeys class option if keyword display desired

\maketitle

YBa${_2}$Cu${_3}$O$_{7-\delta}$ (YBCO) continues to be of great fundamental interest for research into High-\textit{T}$_{c}$ materials. In particular, the study of the magnetic vortex lattice (VL) commands attention~\citep{Mag95,Yet93,Kei94,Bro04,Joh99}, as the VL structure and distortion directly reflect the underlying microscopic state and inter-vortex interaction. In YBCO, superconductivity on the CuO chains, which lie along the crystal \textbf{b}-axis, is believed to be induced by proximity effect from the CuO$_{2}$ planes~\citep{Atk2,Kha06}. This results in an enhanced supercurrent response in this direction relative to the \textbf{a}-axis, which is reflected in the distortion of hexagonal VL structures seen at low fields~\citep{Kei94,Bro04,Joh99}. At low magnetic fields parallel to the \textbf{c}-axis $(\textit{B}\parallel\textbf{c})$, anisotropic London theory relates the value of this distortion directly to the penetration depth anisotropy $\gamma_{ab}$ $(=\lambda_{a}/\lambda_{b}=\sqrt{m_{a}^{\ast}/m_{b}^{\ast}})$~\citep{Cam88,Thi89}. However, other theories also predict that the VL structure and distortion depend on the effects of a nodal gap structure~\citep{Ich99,Ber95,Shi99,Xu96,Aff97}, and on \emph{non-local} electrodynamics combined with a Fermi surface anisotropy~\citep{Kog97}. The common prediction is for the formation of square VL structures at high fields. However, for YBCO specifically, comparison to theory is complicated by the orthorhombic crystal structure and in-plane electronic anisotropy. These will introduce an $s$-wave admixture in phase with the predominantly $d_{x^{2}-y^{2}}$ order parameter, though the exact details remain under debate~\citep{Mull95,Smi05,Kir06,Kha07}. Furthermore, previous studies of the VL in YBCO have been hampered by twin plane pinning along $\{110\}$~\citep{Yet93,Mag95,Kei94,Bro04,Joh99}, which may obscure the intrinsic VL structure. For the measurements reported here, the sample was detwinned and overdoped ($\delta$~=~0) to minimize vortex pinning. As a result, for the first time we have observed VL structures, with $\textit{B}\parallel\textbf{c}$, that avoid the effects of twin plane pinning.

\begin{figure*}
\includegraphics[width=7in]{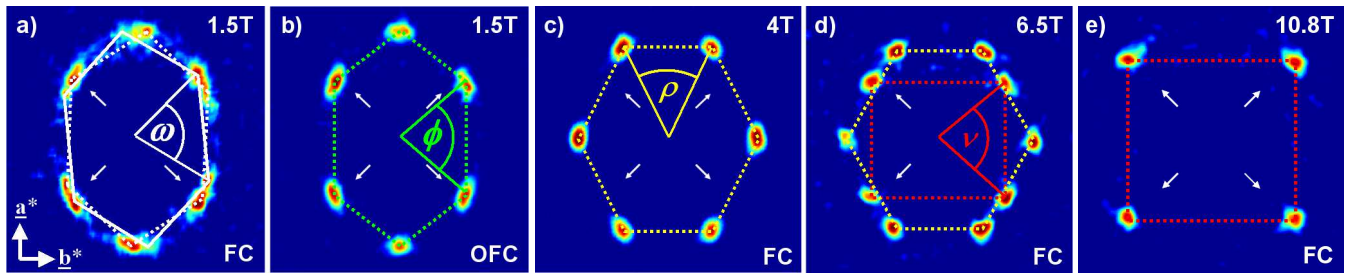}
\caption{(Color online). The VL diffraction patterns taken at 2~K in applied fields,~\textit{B} of (a)~1.5~T after FC, (b)~1.5~T after OFC, (c)~4~T after FC, (d)~6.5~T after FC and (e)~10.8~T after FC. Overlaid patterns indicate the different VL structures that make up the overall diffraction patterns, and the angles between certain Bragg spots. In (b) the angle $\phi$ is bisected by the \textbf{b}$^{\ast}$ direction, in (c) $\rho$ is bisected by \textbf{a}$^{\ast}$ whilst in (d) $\nu$ is bisected by \textbf{b}$^{\ast}$. In all cases, white arrows indicate the $\{110\}$ directions. The diffraction patterns were constructed by summing the detector measurements taken for a series of sample angles about the horizontal and vertical axes. Statistical noise around the main beam at the center of the image has been masked, and the pixelated data has been smoothed with a Gaussian envelope smaller than the instrument resolution. The \emph{real space} VL can be visualized by rotating the reciprocal space image by 90$^{\circ}$ about the field axis and adding an additional spot at the center.}
\end{figure*}

To image the VL structure directly, we have used the bulk probe of small-angle neutron scattering (SANS) in a series of experiments at the ILL and PSI. Neutrons of wavelengths between 6~-~10~\AA~were selected with a 10$\%$ FWHM spread and collimated over 6~-~14~m before the sample. Diffracted neutrons were collected by a position adjustable 2D multi-detector. The sample, of total mass $\approx$~30~mg, was a mosaic of six single crystals, co-aligned to $\lesssim$~1.5$^{\circ}$ about the \textbf{c}-axis. The crystals were grown in a BaZrO$_{3}$ crucible~\citep{Erb96}, detwinned under uniaxial stress at $\approx$~500$^{\circ}$C for 24 hours~\citep{Hin07}, and oxygenated to O$_{7}$ under high pressure~\citep{Erb99}. Using neutron diffraction to observe (100) and (010) reflections from the sample, no minority twin-domain signal was detected at the 1~$\%$ level. The mosaic was mounted with the \textbf{a}-axis vertical and the \textbf{c}-axis parallel to the neutron beam. With \textit{B}$\parallel$\textbf{c}, the VL was prepared using two different techniques. The first was a standard field-cool through $T_{c}$ in a constant amplitude field (``field-cooling'', FC). The second was a field-cool in a field oscillating around the target value (``oscillation field-cooling'', OFC) with an amplitude between 0.1~$\%$ and 0.2~$\%$ of the target field, and a frequency of $\approx$~2~min$^{-1}$. For both FC and OFC, once at the measuring temperature of 2~K, the field was held stationary for diffraction measurements. These were performed by rotating the cryomagnet and sample together to angles that brought various diffraction spots onto the Bragg condition at the detector. In all cases, background measurements were taken at $T$$>$$T_{c}$, and subtracted from 2~K foregrounds, to leave just the VL signal.

Figs.~1(a) to (e) show VL diffraction patterns collected at 2~K and at various fields. The results at 1.5~T (Fig.~1(a) and (b)), reveal the marked difference in VL structure we observe depending on whether the VL at low fields is prepared by FC or OFC. In Fig.~1(a), close inspection of the VL structure reveals it to be composed of two distorted hexagonal domains, each with two of their six spots corresponding to $\{$110$\}$ planes. These structures are indicated in Fig.~1(a). Similar VL structures arise in one of the crystal domains of a twinned crystal~\citep{Bro04}, with the VL orientation controlled by twin planes. In the present case, there was no detectable signal from VLs with a distortion corresponding to the other crystal domain. Nevertheless, we cannot rule out pinning by twin planes associated with a \emph{very small} fraction of minority crystal domains in our sample. However, the pinning potential may instead be due to minor crystal disorder along $\{110\}$ planes remaining after the removal of twins. After OFC at 1.5~T, Fig.~1(b) shows that the VL structure is instead composed of a \emph{single} distorted hexagonal domain. The structure change is also accompanied by a $>50\%$ mean reduction in the rocking curve width of the Bragg spots, and $>20\%$ mean reduction in their azimuthal width. These correspond to substantial improvements in the longitudinal correlation of the vortices in the field direction, and in the orientational order about the field axis respectively. This improvement in VL perfection suggests that the vortex motion caused by OFC allows the VL to explore the true minimum ${F_m}$ of its free energy $F$, at temperatures \emph{below} the FC irreversibility line.  Since anisotropic London theory predicts all orientations to be degenerate~\citep{Thi89}, we expect $F_m$ to be shallow at low fields, and OFC required to reveal it.

At fields from 2.5 to 6~T, with stronger vortex-vortex interaction, the effect of OFC is observed to be less important; whether the VL is prepared by FC or OFC, the VL structure is composed of a single distorted hexagonal domain like that in Fig.~1(c). Note that this intermediate-field structure is rotated by 90$^{\circ}$ relative to that in Fig.~1(b). However, in the field region between 1.5 and 2.5~T, the structures illustrated in Figs.~1(b) and (c) will have very similar -- and even more shallow -- $F_m$, so the free energy $F$ must vary little with VL orientation. Under these circumstances, pinning may again be expected to play a role. Indeed, on FC at 2~T, we observe that the diffracted intensity forms an almost continuous ellipse, indicating VLs pinned in a range of orientations of similar $F$. Even with OFC at 2~T, due to the absence of a dominant intrinsic $F_m$, pinning remains important, as we observe the re-emergence of the \{110\}-pinned VL structure of Fig.~1(a). However, now this structure \emph{co-exists} up to 2.25~T with that illustrated in Fig.~1(c). In no case do we observe a \emph{smooth} transition between the two VL orientations represented by Figs.~1(b) and (c), which indicates that the intrinsic transition between them is likely to be \emph{first-order}. Further evidence for this is given by the field-dependence of the hexagonal structure distortion. For each VL structure, this is characterized by the axial ratio, $\eta$, of the ellipse that overlays the Bragg spots in each domain. The value of $\eta$ may be calculated from the measured angles between VL reciprocal lattice vectors (all angles would be 60$^{\circ}$ for the isotropic case of $\eta~=~1$). Angles (approximately) bisected by symmetry axes (i.e. $\omega$ in Fig.~1(a), $\phi$ in Fig.~1(b) and $\rho$ in Fig.~1(c)) provide the most sensitive measures of $\eta$. The resulting field-dependence of $\eta$ is presented in Fig.~2. We note that for the OFC data, Fig.~2 reveals a clear discontinuity in $\eta$ at $\approx$~2~T, which is further evidence for the first-order nature of the re-orientation transition.

\begin{figure}
\includegraphics[width=3.25in]{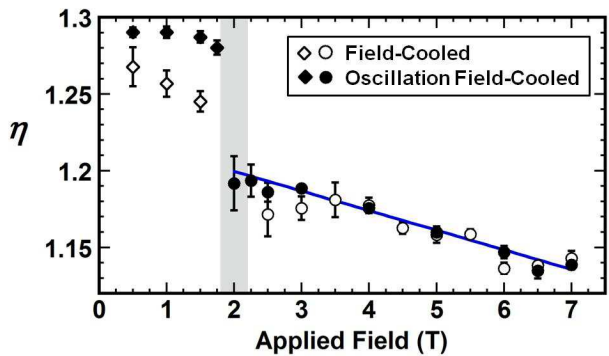}
\caption{(Color online). The field-dependence of the axial ratio, $\eta$ of the ellipse that overlays the Bragg reflections in each VL domain. Triangles, diamonds and circles represent structure types illustrated in Figs.~1(a), (b) and (c) respectively. Open and filled symbols represent VLs prepared by FC and OFC respectively. Error bars (including those in Fig.~3) indicate the standard error. For a VL structure like that in Fig.~1(a), we present the mean value of $\eta$ for the two domains. The blue line is a guide to the eye for the OFC data for the VL structure like that in Fig.~1(c). The shaded portion indicates the field range in which we propose the first-order VL structure transition occurs.}
\end{figure}

Further increase of the field yields another first-order VL structure transition. Fig.~1(d) shows that by 6.5~T the single distorted hexagonal structure like that in Fig.~1(c) co-exists with a newly emerged domain, that we refer to as `rhombic.' Note that for the distorted hexagonal structures, there are six spots of comparable intensity, and only four in the rhombic phase. Remarkably, Fig.~1(d) also reveals that the long axis of the distortion of the rhombic diffraction pattern lies along \textbf{b}$^{\ast}$, opposite to that of the hexagonal structure which lies along \textbf{a}$^{\ast}$. The clear co-existence of the two structures at 6.5~T, and the rapid crossover of diffracted intensity from one domain to the other, indicate the transition to be first-order. At 7.5~T, the VL in the entire sample is composed solely of the rhombic structure, whose aspect ratio falls monotonically towards unity with increasing field. By 10.8~T (Fig.~1(e)), the vortex nearest neighbor directions are close to $\langle110\rangle$, showing a \emph{nearly} square structure.

An inspection of Figs.~1(b)~-~(e) shows that none of the \emph{intrinsic} VL structures have VL planes exactly parallel to $\{110\}$. In all previously investigated samples \citep{Yet93,Kei94,Bro04,Joh99}, one plane in each VL domain was pinned to a $\{110\}$ twin plane, manifest or residual, for any field $\textit{B}\parallel\textbf{c}$. As a consequence, those VL structures evolve \emph{continuously} at high field into the nearly square structure, maintaining the effects of twin plane pinning~\cite{Bro04}. In our study, the weakness of $\{110\}$ pinning allows the intrinsic intermediate-field phase (Fig.~1(c)) to be observed for the first time, and to undergo the expected first-order transition~\citep{Ich99}, to a nearly square phase at high field.

\begin{figure}
\includegraphics[width=3.25in]{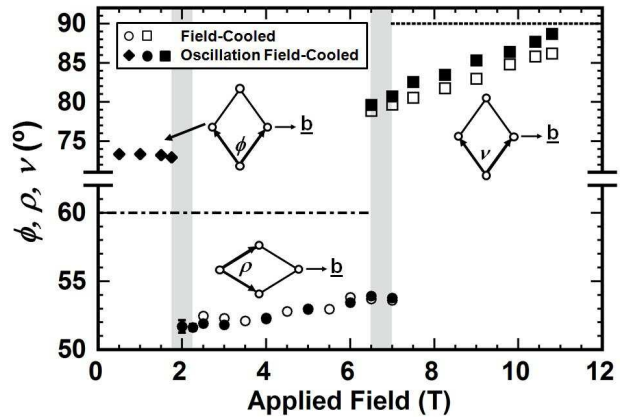}
\caption{The field-dependence of the VL unit cell apex angle for the various structure types. For clarity, we only show measurements for VL structures aligned with the crystal axes, with inset diagrams showing the \emph{real} space orientation of the various VL unit cells. The indicated angles in these diagrams correspond exactly to those shown in the diffraction patterns of Fig.~1. Diamonds, circles and squares indicate VL structures like those in Fig.~1(b), Fig.~1(c) and Fig.~1(e) respectively. Open and filled symbols represent VLs prepared by FC and OFC respectively. Error bars not visible are comparable with the symbol size. The dotted line indicates an apex angle of 90$^{\circ}$ for a perfectly square structure. The dash-dot line indicates an apex angle of 60$^{\circ}$ for an isotropic hexagonal structure. Shaded portions indicate field regions over which we observe first-order VL structure transitions.}
\end{figure}

Fig.~3 presents a more detailed view of the VL structure and distortion. For each structure type, Fig.~3 shows the field-dependence of the apex angle between the basis vectors of the primitive cell. The analysis confirms that, in spite of the effect of OFC which appears to promote a more square structure at high fields, even at maximum field we do not observe a perfectly square VL. It is possible that such a structure occurs beyond our available field range. However, since the crystal is orthorhombic, there is no symmetry reason for the \emph{stabilization} of a square VL structure. Fig.~3 also shows the two shaded field regions over which we observe first-order structure transitions. The size of these regions probably indicate slight sample inhomogeneity, with different structures being frozen-in in different parts of the sample. From these however, we estimate the low-field transition occurs at 2.0(2)~T, and the high-field transition at 6.7(2)~T. Preliminary measurements indicate that these transition fields are not strongly $T$-dependent. Fig.~3 also shows that the rhombic structure phase has more noticeable differences in the precise shape of VLs prepared by FC and OFC, than in the intermediate field hexagonal structure phase. This apparent change in $H$-dependence indicates that the transition at 6.7~T coincides with a change in the mechanism controlling the VL structure alignment.

As noted earlier, anisotropic London theory for high-$\kappa$ superconductors~\citep{Thi89} cannot explain the first-order re-orientation transition that we observe at 2~T. More promising alternatives lie in extensions to the anisotropic London model containing non-local corrections that couple the VL to properties of the Fermi surface. Examples are an $s$-wave theory~\citep{Kog97}, or a similar theory allowing for a $d$-wave gap~\citep{Fra97}. Each predicts first-order 45$^{\circ}$ re-orientation transitions when the field is parallel to a four-fold symmetry direction. In the two-fold symmetry of YBCO, we observe instead a first-order 90$^{\circ}$ re-orientation transition. To overcome the detailed differences between prediction and experiment, it is likely that higher-order terms are required.

At higher fields however, the first-order transition at 6.7~T is more likely to be driven by the $d$-wave order parameter. Notably, the $d$-wave theory of~\citep{Ich99} predicts the transition between hexagonal and square VL structures to be first-order, as their free energies \emph{cross}. Currently available theories~\citep{Ich99,Ber95,Shi99,Xu96,Aff97} treat tetragonal systems with a pure $d$-wave order-parameter, but we expect that YBCO is close enough to this limit that valid comparisons can be made. At high $B$ and low $T$, such theories predict that square structures are stabilized with the vortex nearest neighbor directions lying \emph{along} the nodes of the gap. For a pure $d_{x^{2}-y^{2}}$ order parameter, this is along $\langle110\rangle$, in agreement with the directions approached at high $B$ by the vortex nearest neighbors in the rhombic phase.

Finally, we discuss the detailed shape of the VL structures. Within anisotropic London theory, the hexagonal VL structure distortion parameter, $\eta$ = $\gamma_{ab}$, the penetration depth anisotropy~\citep{Thi89}. On referring to Fig.~2, we note that our values from the OFC data at low fields agree with previous work~\citep{Kei94,Bro04,Joh99}. However, if the structure transition at 2~T is driven by non-local effects, this will render $\eta\neq\gamma_{ab}$, and restrict the local regime only to the lowest fields. Nevertheless, above 2~T, the monotonic fall of $\eta$ with increasing field does suggest the suppression of an underlying anisotropy. This could be related to a possible field-driven suppression of the proximity-effect induced superconductivity on the chains~\citep{Atk2}. Importantly, at 7~T there remains a significant $a$-$b$ anisotropy that is likely to persist to higher fields. Hence, on moving through the transition at 6.7~T, the change in the sign of the distortion between the hexagonal and rhombic structures is unlikely to be described by a Fermi surface anisotropy. Instead we consider the structure of the order parameter. Recent very low field phase-sensitive measurements~\citep{Kir06} show that the $d_{x^{2}-y^{2}}$-like nodes do not lie exactly along the $\langle110\rangle$ directions, but at $\approx$$\pm$50$^{\circ}$ about the \textbf{b}-axis. Such an imbalance in the lobes may be described by a $d + s$-wave order parameter admixture, reflecting additional superconductivity on the chains. Since we expect the vortex nearest neighbors to lie along the nodal directions (e.g.~\citep{Ich99}), the observed \emph{sign} of distortion of the rhombic structure is qualitatively consistent with the phase-sensitive results~\citep{Kir06}. With increasing field, we expect that the chain gap along \textbf{b} is further suppressed, so that the order parameter approaches $d_{x^{2}-y^{2}}$, and the nodes move towards $\langle110\rangle$. Hence our observation of the vortex nearest neighbor directions approaching $\langle110\rangle$ with increasing field is evidence for a field-driven change in the nodal positions. This may be testable by high-field scanning tunneling microscopy measurements.

In conclusion, we have used SANS to observe \emph{two} field-driven \emph{first-order} VL structure transitions in detwinned YBCO. We argue that the low-field transition at 2.0~T is driven by Fermi surface effects, and the high-field transition at 6.7~T is driven by the order parameter anisotropy. The high-field transition is between hexagonal and rhombic structures with distortions of opposite sign, providing evidence for a $d + s$-wave admixture for the order parameter.

The experiments were performed using the D11 instrument at Institut Laue-Langevin, Grenoble, France, and the SANS-I and SANS-II instruments at the Swiss spallation neutron source SINQ, Paul Scherrer Institut, Villigen, Switzerland. We acknowledge financial support by the UK EPSRC, the DFG in the consortium FOR538, the SNF, and from the European Commission under the 6th Framework Programme through the Key Action: Strengthening the European Research Area, Research Infrastructures, Contract No. RII3-CT-2003-505925.

\bibliography{YBCO}

\end{document}